\documentclass[a4paper]{panl}
\usepackage{cite}
\usepackage{wrapfig}
\usepackage{graphicx}
\usepackage{amssymb}
\usepackage{amsfonts}
\usepackage{amsmath}
\usepackage{longtable}
\usepackage{rotating}
\usepackage{lscape}
\usepackage{epsfig}
\usepackage{multirow}

\usepackage{lineno}
\originalTeX
\begin{document}
% Journal sections (see http://pkp.jinr.ru/index.php/PEPAN_LETTERS/about/editorialPolicies#focusAndScope)
%\issuearea{Physics of Elementary Particles and Atomic Nuclei. Theory}
% or in Russian
%\issuearea{ФИЗИКА ЭЛЕМЕНТАРНЫХ ЧАСТИЦ И АТОМНОГО ЯДРА. ТЕОРИЯ}

\title{%Unified Description of
{\bf Elastic Hadron Scattering at High Energies} }
% ($pp$,$p\bar{p}$,$np$,$\pi^{\pm}p$)}
                       %    at Low and High Energies }
%Oscillations (quasi-periodical structure) of elastic hadron scattering
%              at high energies  \\
%              Осцилляции (квази-периодиеская структура) упругого адронного рассеяния при высоких %энергияx}
%This is the title \\ Название статьи}
\maketitle
\authors{{\it O.V.\,Selyugin} \footnote{E-mail: selugin@theor.jinr.ru} }
%S.\,Author$^{b,}$\footnote{E-mail: second.author@email.ru}}
%\setcounter{footnote}{0}
%\authors{О.В.\,Селюгин\footnote{E-mail: selugin@theor.jinr.ru} },
%И.О.\,Второйавтор$^{b,}$\footnote{E-mail: second.author@email.ru(русский вариант)}}
\from{$^{1}$\, {\bf Joint Institute for Nuclear Research, Dubna, Russia}}
%\from{$^{2}$\, БЛТФ, ОИЯИ, Дубна}  %$^{a}$\,
%\from{$^{b}$\,Affiliation 2}
%\from{$^{b}$\,Место работы автора 2}

%%%%%%%%%%%%% line numbering
%\linenumbers
%%%%%%%%%%%%% line numbering

\begin{abstract}
% Russian translation of the abstract
%модели предсказывали что дифракционные взаимодействия адронов перейдут в новый режим  на БАК:
% дающий громадную энергию при которой  S-матрица достигнет унитарного предела \cite{CS-PRL}.
%   \\
%\vspace{0.2cm}
%
  A brief historical overview of various modern approaches to the problem under consideration is given. It includes existing models based on a sum of different terms of the scattering amplitude with different signs and Regge-eikonal models based on the Born terms of the scattering amplitudes.
  %The presence of a periodic structure in differential cross sections at LHC energies
  % As some example of such model
  %is considered by a model-independent method and within the framework of
  % the new Regge-eikonal model
  % taking into account the generalized structure of nucleons (HEGS model) which
  % based on the analyticity of the scattering amplitude.
   An example of such a model is a new Regge-eikonal model is given, taking into account the generalized structure of nucleons (the HEGS model), which is based on the analyticity of the scattering amplitude.
   %, which made it possible
    A unified quantitative  description of various hadron reactions and a  description of differential cross sections and the spin-correlation parameter for interactions    were obtained. % in the frame of the model. % [4]. This allowed us to substantiate the obtained hadron structure from the generalized parton distributions we used.
    In the framework of the model, the existence of experimental  data of elastic hadron
        scattering in the energy range of LHC  and in a wide energy region
    $\sqrt{s}=3.6 -13000$ GeV  was describe a quantitatively
     from a unified point of view.
 The predictions for $\sigma_{tot}(s)$ at superhigh energies are presented.
The possible thin structure of differential cross sections at small angles
of elastic nucleon-nucleon scattering at high energies is discussed.
%
%\vspace{0.5cm}
%%  \hspace{4cm}                             {\bf \large  Аннотация}         \\
%
%  Дан краткий исторический обзор различных современных подходов к рассматриваемой проблеме. Он %  включает существующие модели, основанные на суммировании различных
%членов амплитуды рассеяния с разными знаками, и Редже-эйкональные модели,
%основанные на борновских членах амплитуд рассеяния. В качестве примера такой модели приводится %новая Редже-эйкональная модель, учитывающая обобщенную структуру
%  нуклонов (модель HEGS), основанная на аналитичности амплитуды рассеяния.
%  Получено единое количественное описание % различных адронных реакций, а также описание
%дифференциальных сечений и параметра спиновой корреляции для различных адронных реакций.  %%взаимодействий.
% В рамках модели получено количественное описание % существование
%экспериментальных данных по упругому рассеянию адронов в области энергий LHC и
%в широкой области энергий от $\sqrt{s} = 3.6$ до $\sqrt{s} = 13$  ТэВ
% % √s = 3,6 − 13000 ГэВ
% с единой точки зрения.
%Представление эйконала и амплитуды рассеяния через прицельный параметр позволяет нам %анализировать эффекты насыщения унитарности при сверхвысоких энергиях, включая энергию %выше %LHC.
% Представлены предсказания для  $\sigma_{tot}(s)$ % σtot(s)
% при сверхвысоких энергиях. Обсуждается возможная тонкая структура дифференциальных сечений % упругого рассеяния нуклонов под малыми углами при высоких энергиях.
\end{abstract}
\vspace*{6pt}


%\label{sec:intro}
%\section*{Introduction}

%   \vspace{36pt}

%   \vspace{2cm}
%\label{sec:intro}
\section*{Introduction}

\label{intro}
%  \Russian
%  модели предсказывали что дифракционные взаимодействия адронов
%  \Eng

%   Ryskin \\Определение: Дифракция = упругое рассеяние, вызванное искажением падающей %   волновой функции. \\

    Hadronic high-energy physics has progressed tremendously since the early days of the
S-Matrix theory and Regge poles. Initially, it was believed that the imaginary part
of the (spin-non-flip) amplitude
 can be negligible at asymptotic energies \cite{Eden};
 %\bibitem{Eden} R.J. Eden, "High Energy Collision of Elementary Particles (1967).
 however, the Serpuchov effect \cite{Serp-eff} shows that this
 is  really what   matters in the forward direction
because of the constraints coming from unitarity \cite{mart,roy}. %   [2].
%Первоначально считалось, что мнимая часть
%амплитуды (спин-непереворот) была тем, что действительно имело значение в прямом направлении
%из-за ограничений, вытекающих из унитарности
Eventually, this led to the birth and
growth of the pomeron philosophy. As time went by and the analysis got more and
more refined and higher and higher energies were explored, people realized that the
actual picture was considerably more complex. Analyticity soon showed that one could
not do without a real part \cite{roy} %[3, 4]
 while polarization data proved that it was not possible
to ignore spin complications.
  There are   some problems   as confinement, hadron interaction at larger distances,
  connection of quark with the Higgs field \cite{Petrov-Q},
  nonperturbative hadron structure   (parton distribution functions (PDFs),
 generalized parton distributions (GPDs) and others)
 that  should be explored in the framework of the Standard Model.
  These problems are connected with the hadron interaction at high and super-high energies   and with the problem of  energy dependence of the structure
 of the  scattering amplitude and  total cross sections.

 The research into the structure of the elastic hadron scattering amplitude
  at superhigh energies and small momentum transfer - $t$
   is a very important problem as in this area we should find
   can give a connection   between
   experimental knowledge and  the basic asymptotic theorems
  which are based on  first principles. % \cite{Block-85}.
    It gives   information about  hadron interaction
  at large distances where the perturbative QCD does not work,
  and a new theory  such as, for example, instanton or string theories,
   must be developed.
   High energy experiments can give valuable information
   about the basic properties of the scattering amplitudes.
   However, we do not know the real energy region that will open a new area
   for deep understanding of the hadron interactions. Elliot Leader
   notes: "Although asymptopia may be very far away
 indeed, the path to it is not through a desert
 but through a flourishing region of exciting Physics…”. % (CERN Courier).
  But  to obtain the correct predictions of modern theory at asymptotic energy,
  it is necessary to obtain good descriptions of existing experimental data
  in a wide energy region. The lesson of LHC shows that such a situation is not trivial.
  The first experimental data on elastic proton-proton scattering at $\sqrt{s}= 7 $ TeV
  do not coincide with any existing model predictions (see Fig. 1).
	% -Elliot Leader in CERN Courier
% \end{document}

%High-Energy Particle Diffraction
%30 September 2002
%by Vincenzo Barone and Enrico Predazzi, Springer Verlag 2002, ISBN 3540421076,

 %==================Figs.1 =============================
%       \includegraphics{рисунок}, указав имя файла без расширения.
%\label{sec:figures}
%~\vspace{-1.5cm}
\begin{figure}
\begin{center}
\includegraphics[width=0.4\textwidth] {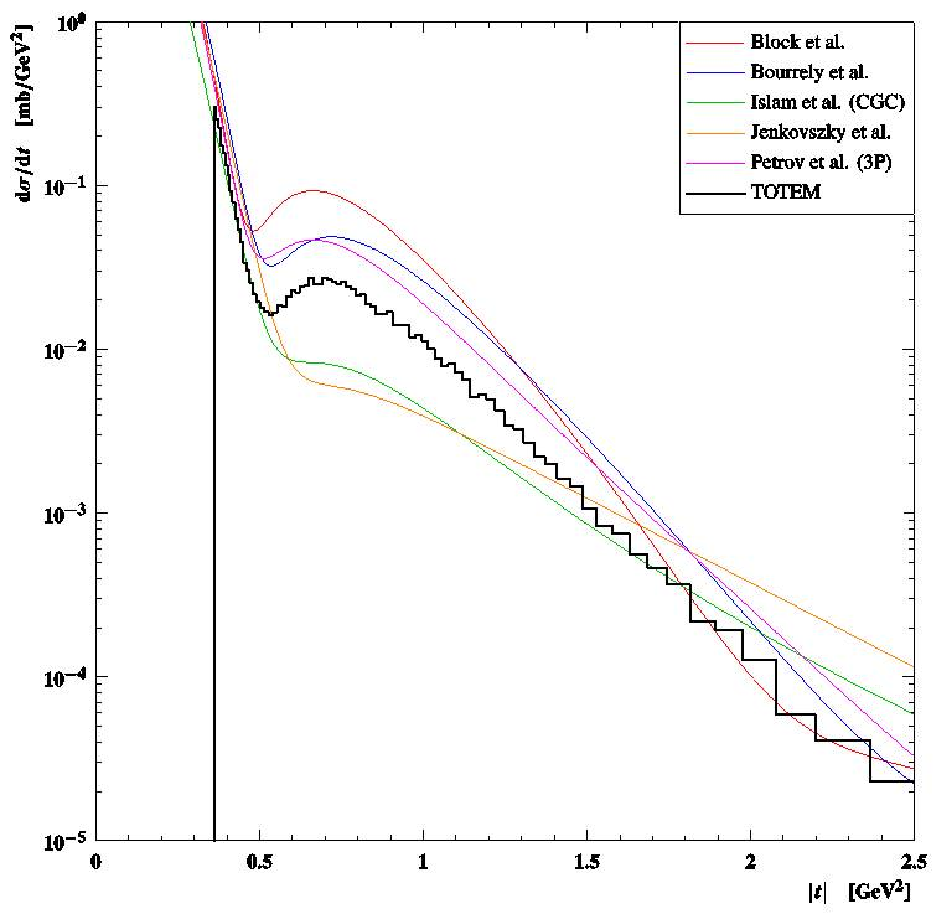}
\end{center}
\caption{Predictions of different models and first data at $\sqrt{s}=7$ TeV at LHC
 \cite{T7ab}.
 }
% (left) and for $p\bar{p}$ (right). }\label{Fig:MV}
\end{figure}

 Modern studies
   of elastic scattering of high energy protons lead to several unexpected results reviewed,     e.g., in \cite{Drem2}.
Some models predict that soft hadron interactions will enter a new regime at the LHC -
%given the huge energy,
 the S-matrix reaches the unitarity limit.
%Analysis of the new effects discovered on the basis of experimental data
%at $\sqrt{s}=13$ TeV \cite{osc13,fd13} and associated with the
%specific properties of the hadron potential at large distances.
%Analysis of
The new effects
discovered on the basis of experimental data at $\sqrt{s} = 13$ TeV    \cite{osc13,fd13} 
 show its association with the specific properties of the hadron potential at large distances.

\section{ Model approximations of hadron elastic scattering }
%\subsection{Diffraction and basic  amplitudes in $t$-channel}

             Diffraction is one of the famous processes of  particle scattering.
   Beginning from  Newton diffraction of a ray on holes which shows that the ray has  wave properties.  At the beginning of the quantum era, the diffraction of particles
   showed that  hard particles also have wave properties.
%   The development of
   The research of
   soft diffraction  at new powerful particle accelerators has opened up
    the “golden age” of Regge theory. Here the emphasis is on those aspects of theory and experiment that are directly relevant to the present-day resurgence of interest in diffraction, i.e., to the diffractive aspects of hard interactions. The essential ideas are dispersion relations  %, Muller’s
    and the generalization of optical theorem to inclusive reactions.
%     and the key, rigorous theorems on permissible growth with energy of cross-sections % are presented.
% Also discussed is the Pomeranchuk theorem relating particle-particle to % %particle-antiparticle asymptotic cross-section growth, but, surprisingly, no attention is
% drawn to the optics-diffraction motivation for the key assumption in the proof of this %“theorem”.
   In the modern language of field theory, we represent diffraction processes as
 the   exchange in the $t$-channel  % of  the colliding particles
 by the certain elementary objects such as
    gamma-quantum, masses mesons, or gluons objects that have  quantum numbers of vacuum - the called crossing even Pomeron, or crossing odd Odderon.
   At high energies, the contributions of the mass second Reggions are dying
   and the main contributions come from the exchange of massless object with vacuum quantum numbers (see Fig. 2).
    Elastic scattering leave s the colliding particles  unchanged. %(see Fig.2).
    In other cases, we obtain  single or double dissociation scattering. Such processes can lead to the creation the Higgs state.
    % (low diagrams     in Fig.3).

%   \end{document}

%      fOR ME \\
%      Figure 37: Diagram for exclusive Higgs production in hadron-hadron collisions. The coupling between
%      This was for instance done by Khoze et al. [408, 409, 410] in an investigation of %      exclusive Higgs production, pp → p + H + p, which has long been proposed as a very %      clean process to detect and study the Higgs at the LHC. In this work the survival %      probability multiplies the result obtained from calculating the diagram in Fig. 37. N %      \\
%      The simplest energy
%      dependence obtained in Regge theory is a sum of “Regge poles”
 %   ThA(s, t) ∼
 The simplest picture considers the Pomeron as two gluon objects which together
    give an object with white color and vacuum numbers. The exchange by three gluons
    gives the crossing odd object - the Odderon.
     Really, quantum chromodynamics can lead to
    many gluon objects with different properties.
     For example, in paper \cite{Kovner} obtained $7$ Pomerons with different intercepts  in the framework of the BFKL approach.
    In the Regge language %the simple pole    gives
    the Pomeron
 %   \end{document}
   with $\alpha_{j}(t)$ is given  as a simple pole – which has the energy dependence
   - $(s/s_{0})^{\alpha}$.
 %  (in $(1 + S_{j}  e^{-i \pi \alpha_{j}(t)} )$.
   $\leftarrow$
 %  and %(326) BFKL Pomeron has $\alpha=0.4$ );
  % douwn here the real-valued coefficients $c_{j}(t)$ depend on the scattering particles,
  % but the “Regge trajectories” $\alpha_{j}(t)$ do not.
  Phenomenologically, the trajectories are well described by a linear form  l
    $\alpha_{j}(t) =  \alpha_{0}(t)+ \alpha^{\prime}_{j}(t) $  
   for not too large $-t$. Each term $j$ is associated with the $t$-channel exchange of definite quantum numbers including the “signature” $S_{j} = \pm 1$
   which distinguishes even and odd terms under   $s \leftrightarrow u$ crossing (equivalent to     $s \leftrightarrow -s$ at high energies).
   In  1984  Landshoff  obtained the properties of the soft Pomeron from the analysis of experimental data  with $\alpha=0.08$ and the hard Pomeron with $\alpha=0.4$.
    %  Notice be pole with the energy dependence  $\ln{s/s0}$;
   % tri
%    a special role is played by exchanges with vacuum quantum numbers ($P = C = 1$)
%   and isospin %ple
%    pole  with the energy dependence  $\ln^{2}{s/s0} +C$
%     $ I = 0$ and positive signature),
%     which dominate elastic amplitudes in the high-energy limit.
%      In QCD these quantum numbers correspond to two-gluon exchange in the $t$-channel.

 %     \end{document}
%       Data for
%      hadron-hadron scattering and for DIS down to $x_{B} \sim 10^{−2}$
%      can be well described by Regge
%      poles corresponding to meson exchange
%      in addition to a “Pomeron” pole with vacuum quantum
%      numbers and a trajectory with $\alpha IP(0) \sim 1.08$ and
%     $\alpha^{\prime} = 0.25$ GeV$^{-2}$  %    GeIP ≈ 0.25 GeV−2 [416, 417].
The HERA data for deep inelastic processes at $x_{B}$ down to about $10^{-4}$
       have however shown that the situation is not as simple, with the rise in energy a slope
           becoming considerably steeper as $Q^{2}$ increases represented
        general properties associated with the phenomenon of diffraction \cite{Frank-4}.

%       The HERA data for deep inelastic processes at $x_{B}$ down to about $10^{−4}$
%       have however shown that the situation is not as simple, with the rise in energy
%           becoming considerably steeper as $Q^{2}$ increases and
%        general properties associated with the phenomenon of diffraction \cite{Frank-4}.
%%    Frankfurt, Strickman, Weiss, and Zhalov (hep-ph/0412260)
% %  The fnalicity require t
%   Analyticity and causality require that the scattering amplitude satisfy the basic %   %requirements of field theory  within axiomatic quantum field theory  such as
%    Froissart – Martin bound $\sigma_{tot} \sim (a_0 + a_1 \ln{s})^2$;
%    derivative dispersion relation $\rho(s,t=0)=\Pi a_1 / (a_0 + a_1 \ln{s})$;
%    Auberson-Kinoshita-Martin scalin $T_{D}(s,t) = i s \ln^2{s} f(|t| \ln{s})$;
%    and crossing properties: $T^{pp}_{D}(s,t) =  T^{p\bar{p}}_{D}(s,t)$

%    \end{document}

  %==================Figs.2 =============================
%       \includegraphics{рисунок}, указав имя файла без расширения.
%\label{sec:figures}
%~\vspace{-1.5cm}
\begin{figure}
\begin{center}
\includegraphics[width=0.8\textwidth] {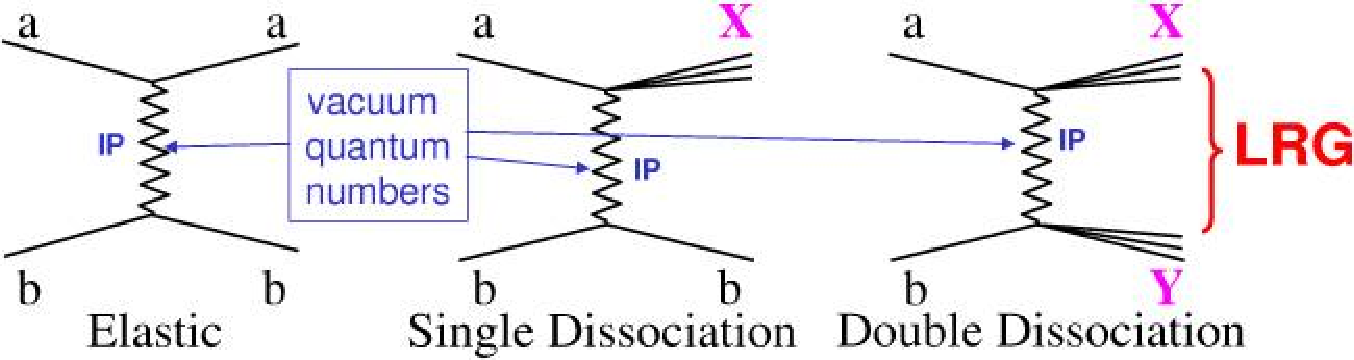}
\end{center}
~\vspace{-1.cm}
\caption{ The elastic hadron scattering
%The ratio of the experimental data of $d\sigma/dt$ of proton-proton elastic  scattering
%   obtained at ISR at  $  \sqrt{s}=19.4$~GeV up to $\sqrt{s}=62$~GeV
%   to the smooth curve (exponential behavior);experimental data %\cite{Barbellini72,Schiz81,Shubert}; (right figure from \cite{Tsarev-2})
%   The fine structure of the diffraction cone at energy $s=50, 500$ GeV$^2$ and   $s=2809, %2016$ GeV$^2$.
% [right] for $pp$ and $p\bar{p}$  at $\sqrt{s}=52.8$ GeV % proton-proton and proton-antiproton
%  scattering.
  % (sqwords, circle- experimental data $pp$ and triangles $p\bar{p}$  correspondingly )% \cite{data}); (lines - the model calculations).
 }
% (left) and for $p\bar{p}$ (right). }\label{Fig:MV}
\end{figure}
%

 %==================Figs.4 =============================
%       \includegraphics{рисунок}, указав имя файла без расширения.
%\label{sec:figures}
%~\vspace{-1.5cm}
%\begin{figure}
%\begin{center}
%\includegraphics[width=0.4\textwidth] {fig3.eps}
%%\includegraphics[width=0.3\textwidth] {fig2.eps}
%\end{center}
%\caption{ % Diagran 2:
%Diagram for exclusive Higgs production in hadron-hadron collisions
% }
%% (left) and for $p\bar{p}$ (right). }\label{Fig:MV}
%\end{figure}
%
%
%
%

\section{%New ISR, SPPS and Tevatron and LHC  data and different models }
% High energy experimental data  and some different models }
  Model descriptions of % experimental
   data on elastic hadron scattering at high energies}

%  There are two main approaches to build the scattering amplitude of hadron elastic %scattering. One cames
There are two main approximations to constructing the elastic hadron scattering amplitude
 at high energies.
The most adequate is based on constructing the Born scattering amplitude and then summing the ladder diagrams using one method or another. The eikonal approximation,
which plays a major role % in later chapters,
when attempting to understand the high-energy behaviour of  %very complex
 QCD Feynman diagrams,
was used in the widely known % Bourrely, Soffer, Wu
 BSF model \cite{Bourrely-1}.

 Other are based on the Landshoff model \cite{Land-2} , which takes into account the first two terms of the eikonal approximation with free coefficients. The result is a set of scattering amplitude terms with different signs,
which allows one to obtain a diffraction pattern with a minimum and a second maximum.
\cite{Land-2a}.
% It can be connected
 This can be due to  gluon connections with different quarks or the same quarks %(see Fig.4)
 which lead to the amplitudes with different signs.
 As a result, they obtained good descriptions of the region of diffraction minimum
and second maximum for % proton-proton and proton-antiproton
  $pp$ and $P\bar{p}$ scattering in the energy range
%from $\sqrt{s}=23$ GeV
up to $\sqrt{s}=546$ GeV.
   This technique allows us to
avoid  complex double calculation of integrals of
complex oscillating functions.
  This was used in Leader's work \cite{GLN}
on which Martynov's model with 40 free parameters was based \cite{Mart-Nicol}.
%  Such approximations was used in the work \cite{Jenk-Sz}
%Laslo Jenkovszky, István Szanyi Modern Physics E Vol. 27, No. 08, 1830005 (2018
%  for the descriptions of the LHC data. % (see Fig.
 In model \cite{Jenk-Sz}, the Pomeron and Odderon contributions were constructed from two terms with different signs, which  %made it possible
  allow to describe the new LHC data.

  As an intermediate option between these two approaches, we can note the work of \cite{FKK}, in which
scattering amplitudes are defined in the impact parameter representation and the eikonal phase is constructed from them.
This work is interesting in that the real part of the amplitude satisfies the local dispersion relation. As a result, the real part has two zeros in the region of small scattering angles.
%E. Ferreira, A.K. Kohara, T. Kodama
%Eur. Phys. J. C (2021) 81:290
%https://doi.org/10.1140/epjc/s10052-021-09056-1
%Analytical representation for amplitudes and differential cross
%section of pp elastic scattering at 13 TeV
%Watanabe \\
  In the modern language,
   the total and differential cross sections of elastic %proton-proton and pion-proton
        $pp$ and $\pi p$ scattering
are studied in the framework of holographic QCD, considering the Pomeron and Reggeon main domain of  elastic scattering as  small angles   in the Regge regime \cite{Watanabe}. % \\

 Some %complicated
 eikonal approximation (two-channel eikonal with three eikonal terms with different coefficients)  was proposed in \cite{Khoze-r24} %,Khoze-r25,Rys-r25}.
 They used a  Pomeron-based formalism incorporating pion-loop insertions in the Pomeron trajectory to account for the nearest singularity imposed by t-channel unitarity and
  also included the Odderon contribution at high energies.
  In \cite{Khoze-r24}, they analyzed the CNI region of $pp$ and $p\bar{p}$
  elastic scattering and obtained a quite satisfactory
  fit with $\chi^2 = 560$ for
504 degrees of freedom. % ν; χ2/ν = 1.11.
 It is interesting that if they neglect the
Odderon contributions, they get a much larger $\chi^2 = 726$.
  %We then extend the framework to include the contribution of an Odderon. P
%[24]
% V. A. Khoze, A. D. Martin, and M. G. Ryskin, Eur. Phys. J. C 18, 167 (2000).
%[25] E. G. S. Luna, V. A. Khoze, A. D. Martin, and M.G. Ryskin,
%Eur.Phys.J.C59,1(2009); E. G. S. Luna, V. A. Khoze, A. D. Martin, and
% M. G. Ryskin, Eur. Phys. J. C 69, 95 (2010).

 %Okorokov \\
%XXXVII International Workshop on High Energy Physics “Diffraction … / Тезисы
%Energy dependence of cross sections in proton-proton and antiproton-
%proton collisions Corresponding Author: vaokorokov@mephi.ru
%(ONLINE)
Note also the works \cite{Ok-2}, where
 the energy dependence of cross sections %, mostly total and elastic,
  was studied for $pp$ and $p\bar{p}$ collisions and continued predictions up to 100 TeV.
 The model with multiple reggeon exchange \cite{OK-P-1} %[2]
 and semiclassical approach with saturation, %[3],
  admit the phenomenon of bosonic condensation.
% Also analytic functions [4] are used
%for approximation of the experimental energy dependence of the ratio of the % %elastic-to-total cross sections in (anti)proton-proton interactions.
%Most of model %dependencies of total cross section on collision energy show the behavior close to the %Froissart–Martin limit in functional sense. On the basis of the resulting approximations, %the total cross section and the ratio of the elastic-to-total cross
%sections for proton-proton collisions are estimated at various collision energies up to an %ultra high values about of 10 PeV in order of magnitude.
They noted that
in general, Bose–Einstein condensation (BEC) can provide a noticeable growth of multiplicity of secondary pions at sufficiently high energies \cite{Ok-2}, %[5]
which can affect the behavior of cross sections. Thus, perhaps, BEC can be suggested as one of the possible dynamical mechanisms which may lead to some novel features in energy dependence of cross sections.

% Ok-P-1,Ok-2    S. D. Campos, V. A. Okorokov, Int. J. Mod. Phys. A 25, 5333 (2010); V. A. Okorokov, S. %D. Campos, ibid 32, 1750175 (2017).

%[2]
%OK-P-1
%V. A. Petrov, V. A. Okorokov, Int. J. Mod. Phys. A 33, 1850077 (2018). [3] V. A. %Okorokov, Phys. At. Nucl. 81, 508 (2018). [4] V. A. Okorokov, Phys. At. Nucl. 82, 134 %(2019). [5] Ok-2 V. A. Okorokov, Phys. At. Nucl. 87, 172 (2024)   \\
%
%

%   S.M. Roy CERN- Geneva Ref. TH.1281-CERN (15.01 1971) \\
% S. M. Roy, Phys. Reports, 5C, 125 (1972)
% S. M. Roy, Phys. Reports 5C, 125 (1972), p.146, Eq.  (4.6b)
%In \cite{Roy-osc,Roy-1}

%\section{
%The elastic nucleon scattering in the framework of the HEGS model
%  in the dip regi
%      HEGS model and fine structure at different energies}
\section{
% ````Model approximation}
          Regge-eikonal High Energy
Generalized Structure (HEGS) model}

      With increasing energy of colliding beams some new effects \cite{CS-PRL}
   in  differential cross sections  can be discovered at small $t$. %\cite{L-range}.
   However,  some new effects have to be analyzed in a wide energy region to check the existence of new phenomena.
   For that, one needs   a model which can describe experimental data in
   a wide energy region.
   We have used our Regge-eikonal high energy generalized structure  (HEGS) model     \cite{HEGS0,HEGS1,HEGS-f}
   that describes the differential cross sections from $\sqrt{s} = 3.6$ GeV up to $ \sqrt{s} = 13.$ TeV.

 %  A simultaneous analysis  is carried out of  90 sets of data.
   In our researches, we use the wide $s$ and $t$ %possible
    region of energy. % experimental data.
   The energy region begins from $\sqrt{s} = 3.5 - 3.8$ GeV for  %proton-antiproton
   $p\bar{p}$ scattering.
    For the  $pp$  %proton-proton
     elastic scattering we took into account sixty five sets of different experiments
    from % the low energy
     $\sqrt{s} = 6.1$ GeV up to the maximal LHC energy $\sqrt{s} = 13$ TeV.
     Summarizing all the sets of %experimental
      data on  elastic scattering at not large angles,
     we took into account  $115$ sets of  different experiments which included $4326$ experimental points on % proton-proton and proton-antiproton
     $pp$ and $p\bar{p}$ elastic scattering.
%     including 235 experimental data on spin correlation parameter $A_{N}(s,t)$
%     at energy $\sqrt{s}=3.6$ GeV up to $\sqrt{s}=25$ GeV.
       We included the data for the spin correlation parameter $A_N(s,t)$ of the polarized %       proton-proton
       $pp$ elastic scattering. This set of data includes 235 experimental data for % %relatively small energies
        $\sqrt{s}= 3.63$ GeV and up to  $\sqrt{s}= 23.4$ GeV. During our fitting procedure, sufficiently good  descriptions were obtained.
For the first time, we also included in our research the elastic proton-neutron experimental data.
     The corresponding data were taken into account beginning from $\sqrt{s} = 4.5$ GeV up to maximal
      energy proton-neutron collisions  $\sqrt{s} = 27.19$ GeV obtained at accelerators.
       On the whole,  we took into account $24$ sets of  experimental data of different experiments
      which supply $526$ of the experimental data. Hence, on the whole, we took into our analysis
       $5027$ experimental data on  elastic nucleon-nucleon scattering.

 %==================Figs.3 =============================
%       \includegraphics{рисунок}, указав имя файла без расширения.
%\label{sec:figures}
~\vspace{-0.5cm}
\begin{figure}
\begin{center}
\includegraphics[width=0.45\textwidth] {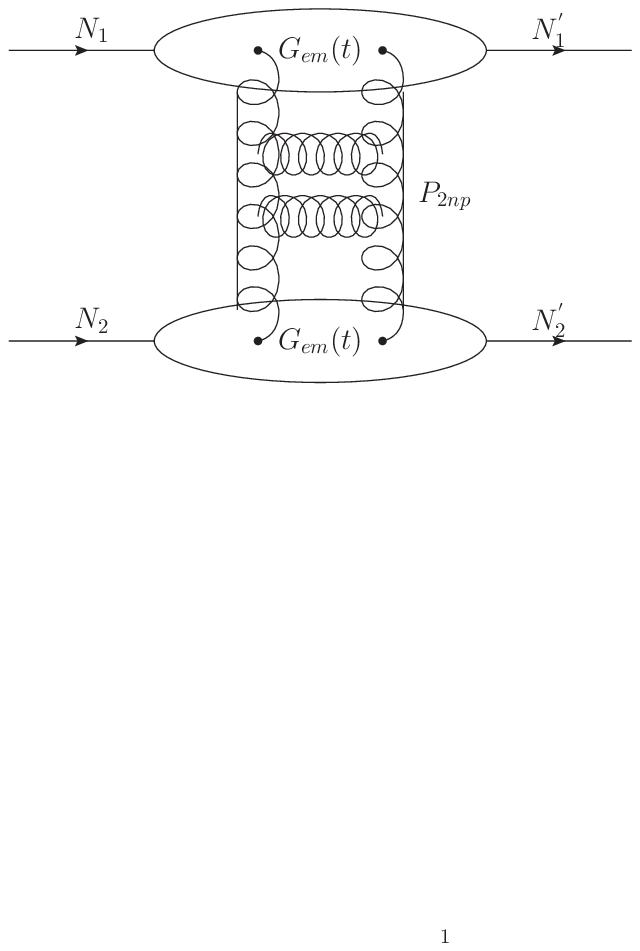} %\hspace{2.cm}
\includegraphics[width=0.45\textwidth] {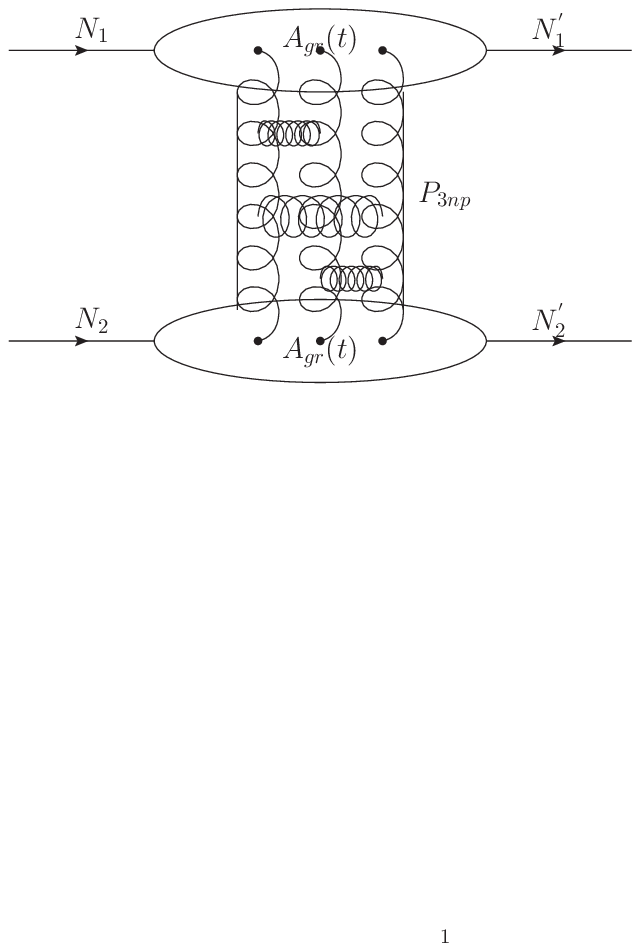}
\end{center}
~\vspace{-4.cm}
\caption{ The Born amplitudes of nucleon-nucleon elastic scattering:
  (left) with two non-perturbative gluons - $P_{2np}$ and (right) three non-perturbative gluons - $P_{3np}$.
	}
% \cite{data}); (lines - the model calculations).
% (left) and for $p\bar{p}$ (right). }\label{Fig:MV}
\end{figure}

 The  model of hadron interaction % is presented that
   is   based on the analyticity of the scattering amplitude taking into account the hadron structure.
    The main features of the model are: a unique energy dependence of the basic asymptotic terms of the Born amplitude
   (all Born terms have one fixed Regge intercept);
the real part of the hadronic elastic scattering amplitude is determined only
   through the complex Mandelstam variable $\hat{S}$ and satisfies the dispersion relations;
   the use of two fixed form factors  determined by different Mellin moments of the same Generalized Parton Distribution (GPDs).
       The calculations of the form factors were
      carried out in \cite{GPD-PRD14}.
   The determination of the $t$ dependence of GPDs up to large valued of $t$ allows us to obtain the inner structure of hadrons and gives the bridge between
   elastic and inelastic reactions.
 %  In the fitting procedure of the experimental data
    The final elastic  hadron scattering amplitude is obtained after the unitarization of the  Born term using the standard eikonal representation.

%~\vspace{-1.5cm}
%==================Figs.8 =============================
%       \includegraphics{рисунок}, указав имя файла без расширения.
%\label{sec:figures}
%~\vspace{-2.5cm}
\begin{figure}
%\vspace{-3.5cm}
\begin{center}
\includegraphics[width=0.8\textwidth] {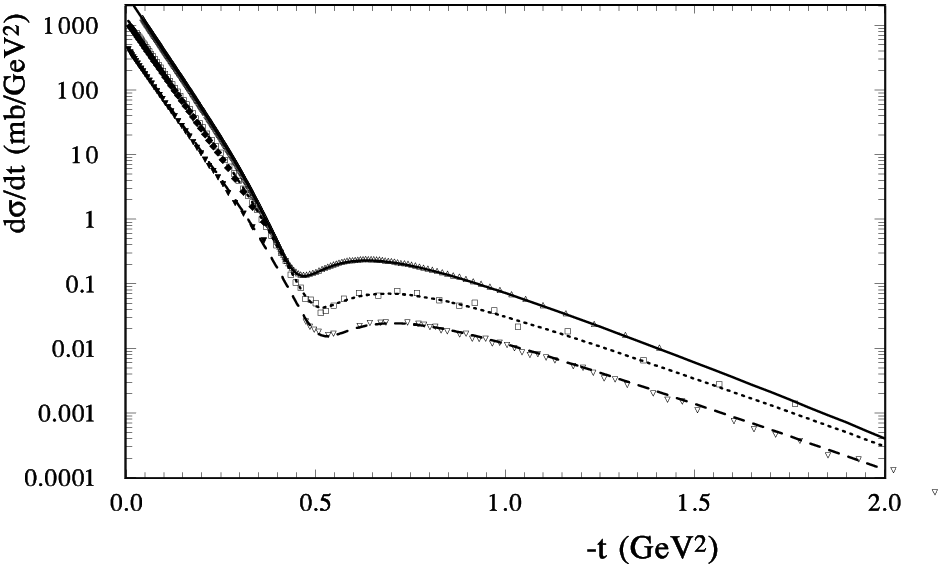}
\end{center}
% ~\vspace{0.5cm}
\caption{The diffraction minimum of  $pp$ %proton proton
 elastic scattering  at  $\sqrt{s}=7, 8, 13$~TeV. }
% ds713m.ps
%
\end{figure}

  The existence of the new effects in high energy elastic $pp$ and $p\bar{p}$
  scattering is revealed by using the data-driving method
for the first time at a quantitative level.
Note that only the presence in the model
of two anomalous terms, determined by the interaction in
the peripheral region allows us to obtain a sufficiently small
$\chi^2/N \sim 1.2$ with $N=4380$.

%%==================Figs. 10 =============================
%%       \includegraphics{рисунок}, указав имя файла без расширения.
%%\label{sec:figures}
%~\vspace{-2.5cm}
%\begin{figure}
%\begin{center}
%\includegraphics[width=0.35\textwidth] {fig12s.eps}
%%\includegraphics[width=0.6\textwidth] {fig8.eps}
%\end{center}
%\caption{Possible odderon contributions calculated in the Khoze-Martin-Ryskin model
%\cite{KMR-18} at $\sqrt{s}=7$ TeV and  $\sqrt{s}=13$ TeV.
%  The hard line - calculations without odderon, dashed line - calculations with odderon
%  contributions. }
%% (left) and for $p\bar{p}$ (right). }\label{Fig:MV}
%\end{figure}

 The crossing odd part of the scattering amplitude, the Odderon, has long been
 a principal goal of high energy particle physics, especially in high energy hadron elastic scattering, and of our understand the properties of QCD at large distances.
%Note,
The  Odderon  arises naturally in perturbative QCD, for example
  %  appears naturally in perturbative QCD
  \cite{KovLevin}, with the intercept $\alpha_{Odd}(0) = 1$ or less.
   %\cite{Bartel        %l00,Kov03}.
%[16, 17].   However,Their wave functions are symmetric under the exchange of
%the transverse coordinates of constituents (the reggeized gluon and the 8S Reggeon).
 The intercept
of this trajectory is equal to that of the BFKL Pomeron.
 It was shown that shadowing corrections decreased Odderon contributions
 and in QCD the C-odd amplitude is expected to be smaller than
the C-even one \cite{Ryskin-Od}.
   % and % they assume that the Odderon gives a contribution of about $1 - 4$ mb at W=7  TeV.  \cite{KovLevin12}.
  %However,
  In the ISR era, the idea of a "Maximal" Odderon \cite{Luk-Nic} which has
  the same intercept as its cross even partner Pomeron, was introduced.
   The experimental data obtained at ISR show the Odderon contribution at ISR energy $\sqrt{s}=52.8$ GeV in the position of the diffraction minimum of $pp$ and $p\bar{p}$ data.
   % and
%   Recently such possibility was confirm
%   at Tevatron energy \cite{Royon} and probably at LHC energy $\sqrt{s}=13$ TeV %\cite{MarNic-Odd}.

%  For example, in \cite{Khoze 1801.07065} it was noted
  %NEW
%  "In other words, in the black disk limit when the value of Re Omega increases
%   and $ exp(−Omega/2) \rightarrow 0$ the Odderon contribution dies out." %  	

% https://doi.org/10.1016/j.physletb.2018.03.025
%       Physics Letters B, 784, 192 (2018)
%"On the other hand, it is possible to introduce the Odderon phenomenologically as an %object which does not violate first principles and the axiomatic theorems.
% In fact it was stated in \cite{NM-rho}
%10] E. Martynov and B. Nicolescu, [arXiv:1711.03288]  Phys. Lett. B 778, 414, 2018
%that the new TOTEM
 result is a definitive confirmation of the experimental discovery of %the Odderon in its maximal form".

%==================Figs.13 =============================
%
%~\vspace{-1.5cm}
\begin{figure}
%\vspace{-1.5cm}
\begin{center}
%\includegraphics[width=0.35\textwidth] {fig12s.eps}
%\hspace{1.cm}
%~\vspace{-1.cm}
\includegraphics [width=0.6\textwidth] {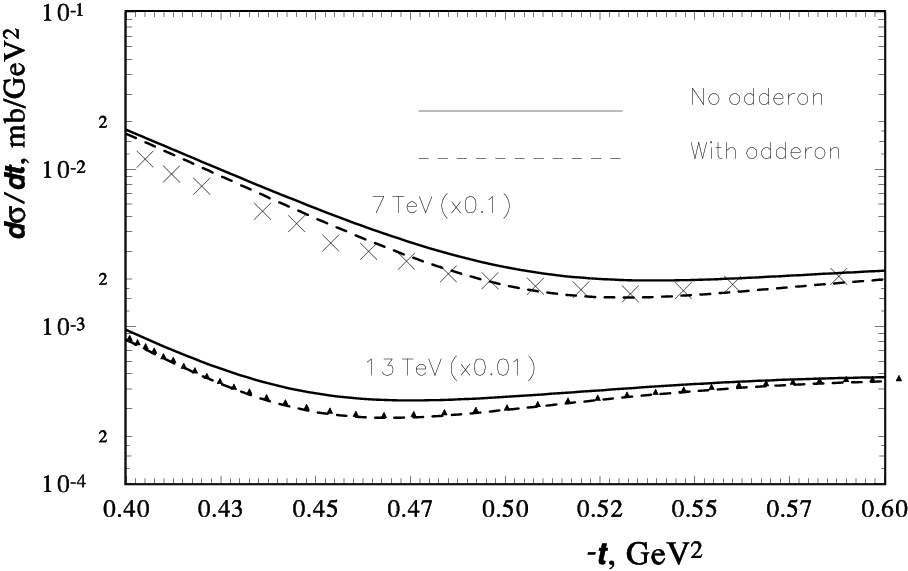} %dm713.eps}
%}
\end{center}
~\vspace{0.5cm}
\caption{ %Possible
The  Odderon contributions at $\sqrt{s}=7$ and $13$ TeV %and  $\sqrt{s}=13$ TeV
 calculated in the HEGS model (
   hard line - calculations without the Odderon, dashed line - % calculations
    with Odderon).}
 % contributions.}
\end{figure}

The presence of the Odderon in our model, with an intercept equal to the Pomeron intercept, leads to a small but clearly pronounced contribution to the differential cross sections in the region of the diffraction minimum (see Fig. 5)
 which drawn in the same form
as in  \cite{Khoze-18}, where the contrary result  was obtained.
%This figure was intentionally drawn in the same manner as Figure 12. It can be seen that %the difference in the cross sections with and without the odderon contribution is %approximately the same in both figures. However, the experimental data at both
% $\sqrt{s}=7$ TeV and $\sqrt{s}=13$ TeV match the curve including the odderon %contribution.
   %   In our fitting procedure % we check up such assumption.
  %  Also   we checked up such a possibility
   We also tested these odderon intercept variants
    and made the fit with two intercepts.
  As a result, we obtained the Odderon intercept $\alpha_{0-Odd}-1=0.1101\pm 0.0004$
  (with fixed Pomeron intercept $\alpha_{0-Pom}-1=0.11$)
  and  that $\chi^2$  did not practically  change.
  Hence, this  confirms our assumption about the equality of  both intercepts.
%     Further
     Many researches are focused on studying the fine structure of the diffraction cone,
     see a short review \cite{Sel-Prot24}.
%=======================
%       \includegraphics{рисунок}, указав имя файла без расширения.
%\label{sec:figures}
%~\vspace{-2.5cm}
\begin{figure}
~\vspace{-1.5cm}
\begin{center}
\includegraphics[width=0.6\textwidth] {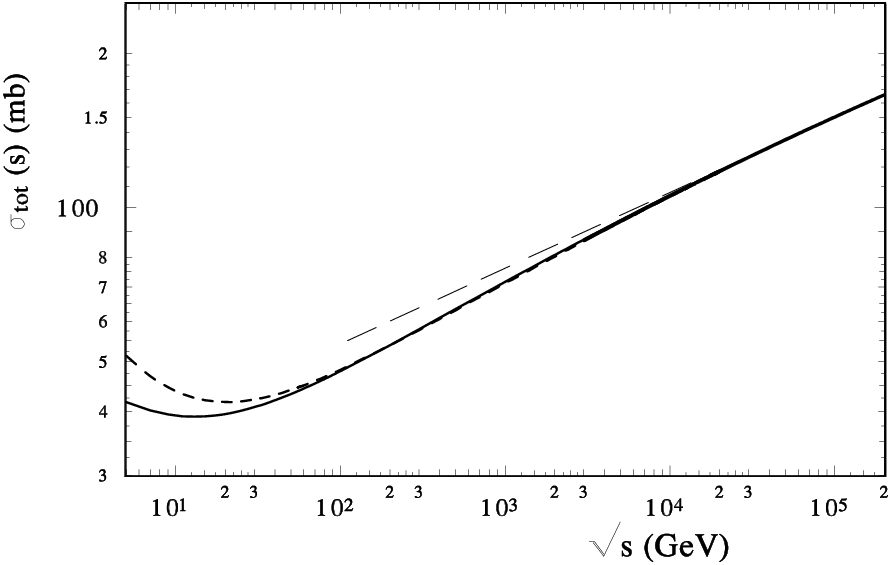}
\end{center}
~\vspace{0.5cm}
\caption{%(left figure) Large $t$  (fig20)
   Total cross section of proton-proton (hard and long dashed lines) and proton-antiproton (short dashed and points
   lines) scattering (thin line shows the approximation of
    the model calculations at super high energies. }
% (left) and for $p\bar{p}$ (right). }\label{Fig:MV}
\end{figure}

%\end{document}

\section{Conclusions}

           New experimental data on elastic proton scattering obtained at the LHC led to significant changes in theoretical models.
           First and foremost, the question of the contribution of the hard Pomeron, with an intercept significantly larger than that of the soft standard Pomeron, was resolved. Its contribution to the differential cross sections of elastic scattering is obviously absent.
           This led to the closure of a number of models and a significant restructuring of others.

           The second important point is the confirmation of the nonlinear behavior of differential cross sections at low momentum transfers but outside the main region of Coulomb-hadron interference. This effect has already been observed in principle in data obtained at the SPS and Tevatron.
           However, convincing evidence of this behavior was obtained at the LHC \cite{T8b}.
           This behavior can be a consequence of many different processes
            such as non-linear Regge trajectory,  contributions of the meson cloud,
%            (J.Pumplin, G.L. Kane; O.V.S.),
            pion loops, % (Anselm, Gribov, Khoze, Martin; Jenkovski et al.),
            different slopes of other contributions
(real part, odderon, spin-flip amplitude), and unitarization.

%\begin{table*} 2
\begin{table}
 \caption{The experimental data of the TOTEM, ATLAS, 
   \cite{HEP-data}
  and HEGS model of the $\sigma_{tot}(s)$, mb and $\rho(t=0,s)$ }
\label{Table-2}
%\vspace{.1cm}
\begin{center}
\begin{tabular}{|c|c|c|c|} \hline
%          &               &                  \\
 $\sqrt{s}$, TeV &$\sigma$   & $\sigma_{tot-exp.}$ & $\rho(t=0,s)$  \\  \hline
 %        &        &       &             \\  \hline
$3$  & $86.5\pm2.9$    & $\pm0.5$  & $ 0.122$  \\
$7$  & $99.8\pm2.9$    & $97.16\pm0.5$ T & $ 0.119$  \\
$7$  & $99.8\pm 2.$    & $97.16\pm0.5$ A  & $ 0.119$  \\
$8$  & $101.7\pm2.9 $    & $99.4\pm0.5$  &$ 0.12$  \\
$8$  & $96.07\pm1.34 $    & $99.4\pm0.5$  &$ 0.12$  \\
$13$ & $110.3 $   & $110.4\pm0.5$ &$ 0.115$  \\
$14$ & $ $   & $108.5\pm0.5$ &$ 0.117$  \\
$25$ & $122.\pm 15$    & $122.7\pm0.5$  &$ 0.112$  \\
$50$ & $136 \pm 23 $    & $$  &$ 0.108$  \\
$100$ & $152 \pm 23 $    & $$  &$ 0.105$  \\
$150$ & $160 \pm 23 $    & $$  &$ 0.102$  \\
$200$ & $166.4 \pm 23 $    & $$  &$ 0.101$  \\
%  &               &                   \\
 \hline
\end{tabular}
\end{center}
\end{table}
%  \end{table*}
          % We note that
            In our model we also found an additional scattering amplitude term at LHC energies with an unusually large slope, which was previously observed only in secondary Regions, whose contribution definitely disappears at LHC energies. Such a term naturally leads to additional nonlinearity of the differential cross sections.
             %The term with large a slope is required for a quantitative description of  %experimental data.
            % in the case of the standard normalization of 13 TeV data.
            The analysis of the whole sets
           of high energy experimental data supports  the existence of such an anomalous term.
           The resulting description of experimental data over a broad energy range will allow for fairly convincing predictions at significantly higher energies. The table presents predictions for the 
           %total and rho
           $\sigma_{tot}$ and $\rho(0)$ 
           % ​​up to an energy of 
          values up to energy of  $200$ TeV. 
           
           Fig. 6 presents the behavior of the $\sigma_{tot}(s)$  %total cross sections
           and shows that the growth of the $\sigma_{tot}$ % total cross sections
           begins to slow at energies above the LHC.

           Unfortunately, the question of the existence of a maximal odderon with an intercept equal to the Pomeron intercept remains open. This remains an important problem  for the study of QCD in the nonperturbative region.
           The question of the actual behavior of the real part of the scattering amplitude as a function of momentum transfer, which must correspond to local dispersion relations, also remains unresolved.
           This question remains model-dependent.

           The long story of research  into the fine structure of the differential cross sections of elastic
            hadron-hadron scattering  \cite{Sel-Prot24} deserves further% study
             and more thorough study.
          %  Up to now
          We still  cannot give a final conclusion about the existence
            of some periodic structure in the elastic scattering amplitude.
             However, many things indicate that such a fine structure exists
             and is determined by the potential of hadronic interaction at large distances.
             % Of course,
             We should   especially %it is need
              note the recent high statistical data
              of the TOTEM Collaboration  at $\sqrt{s} = 13$ TeV  \cite{TOTEM-13}.
             Obvious oscillations were discovered in these data in \cite{osc13}
              using the HEGS model \cite{HEGS0,HEGS1} which describes well the elastic cross section
              of different hadron-hadron reactions simultaneously from   $\sqrt{s} = 3.6 $ GeV
              up to  $\sqrt{s} = 13000$ GeV.  The oscillation picture was confirmed in \cite{Graf-1, Graf-2,Graf-3}
          %    on base a Phillips–
                 using  the Phillips–Barger Regge parametrization \cite{Phil-Barg} plus an oscillating term.
 %                Barger Regge parameterization [14] plus an oscillating term.
                Hence, such an oscillation picture  depends weakly on the basic amplitude
                if the amplitude has  a sufficiently hard form.
                    Contrary, if the basic amplitude has many free parameters \cite{Pet-Tkach-osc} and
                    its form is not hardly determined,
                    a possible oscillatory process manifests itself as
 artificial.
%                    at a level.

%                     not exceeding the experimental statistical error.
%                  We should also emphasize
%                  the recent results of the HEGS model \cite{HEGS-f}.
%                 It was shown that the constant of the oscillation term is determined
%                 with a small error $h_{os}=0.28\pm 0.01$ on the whole experimental data, %                 includeding $4380$
%                 experimental data. The energy dependence of such an oscillation term
%                  is determined substantially well, the main part is proportional to %                  $ln^2(s)$;
%                  hence it has the same energy dependence as the main amplitude.
%                %  Of course, it is require
                Of course, this requires further theoretical research,
                 especially in connection with the long-range potential
                 and new highly accurate experimental data not only at ultra-high energies
                  but also at fairly low energies, which we expect to obtain at the new NICA complex.

%\begin{table}
%\vspace{1.5cm}
%\caption{The non-exponential behavior of the differential cross section at $\sqrt{s}=540$ %GeV and $\sqrt{s}=1800$ GeV }
%\begin{tabular}{|c|c|c||c|c|} \hline
%{\em $ -t $} &
%\multicolumn{2}{c|}{$ \sqrt{s} = 541 GeV $} & \multicolumn{2}{c|}{$ \sqrt{s} =
%1800 GeV $} \\ \cline{2-5}
%$ GeV^{2}$ & \  $ \rho( s,t)  $ \  & $B(s,t) GeV^{-2}$  & $ \; \rho (s,t) \; $
%& $ B(s,t) GeV^{-2}$ \\ \hline
%.001  &   .141      &    16.8    &    .182    &    18.1 \\
%.014  &   .135      &    16.5    &    .178    &    17.7  \\
%.066  &   .112      &    15.5    &    .161    &    16.6  \\
%.120  &   .089      &    14.9    &    .143    &    15.9   \\ \hline
%\end{tabular}
%\end{table}
%%\vspace{1.5cm}
%
%
%
%
%
%
%
% %\vspace*{5mm}
% %
%
\vspace{1.cm}
 {\it Acknowledgements.}
       The author % expresses
        is grateful to the Organizing Committee of the
        XXXVII International Workshop on High Energy Physics
“Diffraction of hadrons: Experiment, Theory, Phenomenology” (2025).
%July 22-24, 2025.
       for the invitation to present the talk that formed the basis of this work.
% The author  would like to thank
%  the organization commetet of
%  the Department of Theoretical and Mathematical Physics of the University of Liege for
% their hospitality and the FRNS for financial support.

\end{document}